\documentclass[conference]{IEEEtran}
\pdfoutput=1

\usepackage{booktabs} 

\usepackage{amsmath,amssymb,amsfonts}
\usepackage{algorithmic}
\usepackage{graphicx}
\usepackage{indentfirst}
\usepackage{enumerate}
\usepackage{multirow}
\usepackage{tikz}
\usepackage{paralist}
\usepackage{listings}
\usepackage{caption}
\usepackage[T1]{fontenc}
\usepackage{balance}

\usepackage[hyphens]{url}

\usepackage{dcolumn}
\newcolumntype{d}[1]{D{.}{.}{#1}}

\usepackage{xspace}

\let\labelindent\relax
\usepackage{enumitem} 

\newcommand{\subhead}[1]{\vspace {1pt}\noindent{\textbf{#1.}}}
\newcommand{\bfdot}[1]{\textbf{ #1.}}

\newcommand{\sol}{Inuksuk\xspace}
\newcommand{\ttsol}{\texttt{Inuksuk}\xspace}

\begin{document}
\author{%
Lianying Zhao and Mohammad Mannan\\
Concordia University, Montreal, Canada\\
\{z\_lianyi, mmannan\}@ciise.concordia.ca\\
}
%
\title{TEE-aided Write Protection Against \\ Privileged Data Tampering}

\maketitle

\begin{abstract}
Unauthorized data alteration has been a long-standing threat since the 
emergence of malware. System and application 
software can be reinstalled and hardware can be replaced, but user 
data is priceless in many cases. Especially in recent years, ransomware 
has become high-impact due to its direct monetization model.
State-of-the-art defenses are mostly based on known signature or 
behavior analysis, and more importantly, require an uncompromised 
OS kernel. However, malware with the highest software privileges 
has shown its obvious existence. 

We propose to move from current detection/recovery based 
mechanisms to data loss prevention, where the focus is on 
armoring data instead of counteracting malware. 
Our solution, \ttsol, relies on today's Trusted Execution 
Environments (TEEs), as available both on the CPU and storage device,
to achieve \emph{programmable write protection}. We back up 
a copy of user-selected files as write-protected \emph{at all times},
and subsequent updates are written as new versions securely 
through TEE. We implement \sol on Windows 7 and 10, and Linux (Ubuntu); 
our core design is OS and application agnostic, and incurs no 
run-time performance penalty for applications. File transfer disruption
can be eliminated or alleviated through access modes and customizable
update policies (e.g., interval, granularity). For \sol's adoptability 
in modern OSes, we have also ported \emph{Flicker} (EuroSys 2008), a defacto 
standard tool for in-OS privileged TEE management, to the latest 64-bit Windows.

\end{abstract}


\section{Introduction and motivation}\label{sec:intro}
The first known crypto-ransomware dates back to 1989 (only file/directory names
were encrypted~\cite{ransom-89-ciac}; see also~\cite{ransom-89-news}).
Crypto-based attack vectors were formally introduced by Young and Yung in
1996~\cite{moti-96} (see also~\cite{moti-17}). After the CryptoLocker attack in
2013, robust crypto-ransomware families have been growing steadily, with a large
number of attacks in 2016 (see the F-Secure ransomware
``tube-map''~\cite{ransom-tubemap}). Examples of recent high-impact
ransomware attacks, include~\cite{wcry, erebus, hospitals, cryptxxx,
cryptowall, lechiffre}, affecting individuals and enterprise/government systems
alike. 
An IBM X-Factor survey of 600 business
leaders and 1021 consumers in the US reveals the effectiveness of current
ransomware attacks: 70\% of affected businesses paid the ransom (46\% of
businesses reported to have been infected); individual users are
less willing to pay (e.g., 39\% users without children may pay ransom for family photos vs.\ 55\% users with children). 
For a conservative estimate of financial loss, a recent end-to-end 
measurement~\cite{tracking} shows that over \$16 million USD in ransoms 
has been collected from 19,750 potential victims over a two-year period.
Ransomware's direct
monetization has benefited from pseudo-anonymous payment systems such as
Paysafecard.com, prepaid/gift cards, and crypto currencies (e.g.,
Bitcoin), and not-easily-traced indirect payments such as sending SMSes to
premium numbers~\cite{gordianknot}. 

Common anti-malware approaches
relying on binary signatures are largely ineffective against ransomware (see
e.g.,~\cite{cryptodrop}).  Some solutions rely on system/user behavior
signatures, exemplified by file system activity monitoring,
e.g.,~\cite{redemption, cryptodrop, shieldfs, unveil}.  To complement detection
based solutions (or assuming they may be bypassed), recovery-based mechanisms
may also be deployed, e.g., Paybreak~\cite{paybreak} stores
(suspected) file encryption keys on-the-fly, right after generated but before
encrypted with the ransomware's public key. 

On the other hand, general (rootkit-level) malware that targets to corrupt/delete 
user data for various purposes also has long existed, e.g., \emph{wiper malware}~\cite{wiper-malware}, without
demanding a ransom, which is worse in terms of recovery. In addition to recent
incidents (e.g., \cite{ohio}), this is also exemplified by the notorious virus CIH~\cite{cih} (1998), which 
erased files and even corrupted BIOS. Several countermeasures against
generic rootkit attacks have also been proposed, focusing on intrusion-resiliency
and forensics (e.g., S4~\cite{selfsecuring}), and preventing persistent infection
(e.g., RRD~\cite{rrd}). FlashGuard~\cite{flashguard} is the only proposal
focusing on rootkit ransomware, which leverages the out-of-place write feature of
modern SSDs, providing an implicit backup. It requires modifying SSD firmware
and a trusted clock within the SSD (currently unavailable). We discuss academic
proposals in more detail in Section~\ref{sec:rel}.

Inspired by traditional data access control (e.g., permission-based
and read-only protection in file systems\footnote{Note: read-only folders/files 
enforced by the OS (e.g., Windows 10 ``controlled folder access''~\cite{win10-folder}), 
only prevents unprivileged access.}) and backup mechanisms, we shift the 
focus from detection/recovery to \emph{data loss prevention} 
against rootkit malware (including ransomware). If user files could remain
unmodifiable by malware even after the system compromise, no reactive
defense would be necessary---enabling data loss prevention. However,
this new paradigm requires that writes must 
be allowed at certain times when new data is added or changes are made. 
Of course, rootkit malware can also make malicious changes to the protected files once
write access is enabled. 

We propose to achieve data loss prevention with an append-only history-preserving 
backup storage framework, where protected files are always exposed
read-only to applications and the operating system, and append operations are allowed
only in a secure manner. In our threat model with rootkit malware, 
such read-only access (equivalent to write protection) need to be enforced by 
hardware/firmware to counteract privileged software. Considering that there exists no 
commercially-off-the-shelf (COTS) storage device with this append-only feature, 
we resort to trusted execution environments (TEEs, see Section~\ref{sec:bg}) to 
allow programmable write protection. We use a host-side TEE to enforce the (append-only) 
logic and confidentiality-protect an authorization secret, and a device-side TEE to allow write access only with that secret.
In a broader sense, the placement of secrets and logic across multiple TEEs relies on
the underlying assumptions and support from available hardware (refer to 
Section~\ref{sec:design} for further discussion). 

By choosing the self-encrypting drive (SED, see Section~\ref{sec:bg}) as the device-side TEE, 
we design \ttsol,\footnote{\sol is an Inuit word with multiple meanings,
including: a (food) storage point/marker.}  
to protect existing user files from being deleted or encrypted by malware. 
\sol functions as a \emph{secure data vault}: user-selected files are
copied to a write-protected SED partition, and the secret to allow write-access
is cryptographically \emph{sealed} to the machine state (i.e., the genuine 
\sol and the correct hardware platform), 
and hence, allowing file writes to the data vault only from the trusted environment.
Deletion/modification outside the environment will fail due to the
write protection.  
Meanwhile, access to the read/writable copy on the original partition 
is not affected (processed at the next commit). \sol takes rootkit 
ransomware as a major threat but also works against any privileged 
unauthorized data alteration.

\sol works without prompting for any user secret. It merely appends to
existing data. Files created or modified on the original partition 
are all treated the same way, and copied (i.e., \emph{committed}) 
onto the protected partition as new \emph{versions} in the host-side TEE 
(referred to as TEE thereinafter if not otherwise specified), without 
overwriting existing files. However, user consent is needed for
solicited file deletion, e.g., when the user no longer needs a document.
Our assumption is that deletions are done occasionally and preferably 
in batch (disk space is relatively cheap). \sol comes with a built-in
mini file browser for the user to select and delete files in TEE.
Files on the protected partition remain accessible as read-only, 
allowing the user to mount the drive elsewhere for recovery without TEE
in the case of malware detection or system corruption 
(see Sections~\ref{sec:design} and \ref{sec:sec}).

We choose to instantiate the host-side TEE using Trusted Platform Module (TPM) chips, 
and CPUs with Intel TXT or AMD SVM (see Sections~\ref{sec:multitee} and \ref{sec:choices} for reasons, and Section~\ref{sec:bg} for background). Due to the exclusive nature (which is also a great security benefit)
of the TXT/SVM environment, during file operations on the protected partition, the
system is unavailable for regular use. We thus provide two access modes for both
home and enterprise users: 
\begin{inparaenum}[a)]
\item \emph{Network-based}. User devices and computers are connected to a regular network storage system; a dedicated \sol computer then copies user data from the regular storage to its protected storage, and thus user experience is not affected. 
\item \emph{Stand-alone}. The user works with the \sol-equipped device, e.g., a laptop. In this mode, system unavailability is reduced with custom scheduling policies (e.g., triggered during idle periods, akin to Windows updates).
\end{inparaenum} 

While \sol can provide strong security guarantees, its implementation faces
several technical challenges.  For example, the TXT/SVM environment
lacks run-time support and we must directly communicate with the SED device (for
security) and parse the file system therein (involving performance
considerations). Note that the use of Intel SGX is
infeasible for \sol, as SGX allows only ring-3 instructions, i.e., cannot access
the disk without the underlying (untrusted) OS.  Also, the user OS is unaware of the TXT/SVM
sessions, so the devices (i.e., keyboard/display for secure user interface)
are left in an unexpected state (see Section~\ref{sec:impl}). 
Note that there have been a series of attacks based on SMM (System Management Mode)
over the past few years, some even affecting Intel TXT~\cite{attack-txt}.
This does not pose a serious threat to \sol, because of its particular setting,
e.g., exclusiveness with no bootloader/hypervisor/OS involved
(discussed more in Section~\ref{sec:sec}).
Last but not least, TXT's exclusiveness also protects it from many side-channel attacks 
that are highly effective against non-exclusive TEEs such as Intel SGX and ARM TrustZone. 

\subhead{Contributions}
\vspace{-4pt}
\begin{enumerate}[leftmargin=1.5em]
\item We design and implement \sol against root-privileged data tampering, in a radical
shift in threat model from existing academic/industry solutions. We target
\emph{loss prevention} of existing data, instead of detection/prevention of
malware/ransomware.
\item \sol's design is tied to the combination of established and standardized 
trusted execution environments
(in our prototype, SED disks and Intel TXT/AMD SVM
with the TPM chip). Integrating TXT/SVM, TPM, and SED/Opal together in a seamless way 
with a regular OS (Windows/Linux) is non-trivial, but offers a significant leap in the 
arms-race against malware. Our solution, together with the ported Flicker (which will be 
both open-sourced), solves several engineering/performance problems when faced with 
exclusive TEEs (e.g., DMA with TEE, disk/file access, display), which can
also be methodologically useful for other TEE applications. 
\item We implement \sol on both Windows and Linux (Ubuntu). The core design is
OS-agnostic. Our prototype achieves decent disk access performance within 
the OS-less TXT/SVM environment (around 32MB/s read and 42MB/s write), when committing 
files to the protected partition. The regular disk access to original files from the 
user OS remains unaffected, i.e., all applications perform as before.
\item We also port the state-of-the-art \emph{in-OS} trusted execution manager 
Flicker~\cite{flicker} to Windows 10 64-bit (Flicker's latest version only supported Windows 7 32-bit).
This advances privileged trusted execution (as opposed to user-space only) up to date,
available to other secure processing applications in modern 64-bit operating systems.
\item 
Beyond unwanted modifications of protected data, \sol can be used as a generic 
secure storage with fine-grained access control, enabling read/write operations 
and data encryption (with \sol-stored keys), if desired. \sol in the stand-alone
mode is locally enforced without any network dependency, and operates with a small TCB.
\end{enumerate}

\section{Background}\label{sec:bg}
In this section, we briefly explain certain terms and background information
to facilitate understanding of the \sol design and prototype implementation
hereinafter.

\subhead{Trusted Execution Environment (TEE)}
Modern CPUs usually support a special secure mode of execution, which
ensures that only pre-configured unaltered code can be executed, with
integrity, secrecy and attestability; and provides a form of isolation
from both other software/firmware and physical tampering. TEEs can be
\emph{exclusive}, preempting and suspending other code (e.g., Intel TXT and AMD SVM), 
or \emph{concurrent}, co-existing with other processes (e.g., Intel SGX, ARM
TrustZone, and AMD SEV~\cite{sev-sgx}). There are also \emph{privileged} (TXT, SVM, SEV and TrustZone) 
TEEs and \emph{unprivileged} TEEs (SGX). This is about whether
privileged instructions (e.g., I/O) are allowed or untrusted OS has to 
be relied on to provide such services. 

Technically, TEEs cannot function alone. For the purpose of storing
measurements (to be matched with that of the code being loaded) and
secure storage of execution secrets, a Secure Element (SE) is used in 
conjunction. It can be part of the processor die, an integrated chip,
or a discrete module.

In this paper, we use the term TEE in a broader sense,
i.e., referring to the aforementioned secure processing feature
of any processor-equipped devices that operates with secrets, 
in addition to that of PC and mobile platforms. For instance, more
and more IoT devices make use of microcontrollers with ARM 
TrustZone for Cortex-M such as Nordic nRF91~\cite{nordic} and
NuMicro M2351~\cite{nuvoton}. Also, 
there are legacy devices with secure processors (cf.~smart cards~\cite{sc}).

\subhead{Intel TXT and AMD SVM}
Trust Execution Technology (TXT) is Intel's first ``late launch'' technique,
aiming at establishing trusted execution any time after system reboot,
without relying on what has already been loaded (e.g., BIOS). It is
exclusive, removing software side-channel attack vectors and with the
help of VT-d~\cite{vtd}, largely defends against violations from 
the I/O space.
AMD SVM (Secure Virtual Machine) is a similar technology, which we consider as equivalent to TXT,
with slight differences, e.g., it does not involve an
explicit ACM (Authorized Code Module, such as TXT's SINIT) and has
fewer requirements for the measured program (called SLB in SVM and MLE in TXT).
Nevertheless, they share the security properties we need for \sol
so we refer to them undistinguished as TXT/SVM hereinafter.
When individual terms are used, the discussion is specific to one.

Note that TXT/SVM has different positioning than Intel SGX and can 
handle privileged instructions, e.g., device I/O as needed by \sol.
They do not replace each other. TXT is widely supported by many
commercial-off-the-shelf Intel CPUs,\footnote{47 CPUs released in 2018 as of Dec.\ 6
(ark.intel.com).}
and SVM (SKINIT) is available in almost all modern AMD CPUs. \looseness=-1

\subhead{TPM}
Trusted Platform Module is a microchip, serving as the SE for TEEs 
(usually TXT or SVM). Its volatile secure storage includes PCRs 
(Platform Configuration Registers) where the run-time measurement 
can be stored and matched with. They can not be directly accessed 
but only extended (i.e., replaced with the cryptographic hash value
of its original value concatenated with the new measurement). 
Its non-volatile secure storage is called NVRAM, which is accessible 
in the form of index (a numeric identifier). NVRAM indices can be 
allocated and deallocated and there can be multiple of them.

\subhead{Sealing}
Short for cryptographic sealing, it is a special mode of encryption,
provided by TEEs/SE, 
where the key is tied (in various ways) largely to the machine state, in the 
form of \emph{measurement}. Measurement is the chaining of the loaded
programs in sequence, e.g., concatenation of hashed values in the SE.
Any single bit of change in loaded programs will cause a mismatch of measurement,
making the reproduced key different, and thus render the decryption (unsealing)
to fail. In this way, platform binding is achieved.

\subhead{Flicker~\cite{flicker}}
Before the introduction of Flicker, Intel TXT was mostly applied with the pilot
project tboot, which deals with boot-time trusted execution (cf.\ OSLO~\cite{oslo} for AMD SVM). 
The ability to switch between the regular OS environment and the trusted execution
was not available. Flicker enables such transitions, e.g., interrupting
and saving states for the OS, initiating the TXT/SVM session, performing
trusted operations and resuming the OS. The trusted operations are
encapsulated in what is called a PAL (Piece of Application Logic) and
thus OS-agnostic. It satisfies what is needed in \sol. 

\subhead{Self-Encrypting Drive}
With the same interface and form factor, regular hard drives or SSDs
can be equipped with a built-in crypto engine and certain enhancement
to the controller, thus providing on-device encryption and access 
control. Such devices are called Self-Encrypting Drives (SEDs)~\cite{sedhw}.
Instead of more generic secure processing, functionalities are mainly
related to media access control and data encryption (e.g., with the so-called 
Device Encryption Key).
Most SEDs offer fine-grained protection, such as dividing media space
into ranges and splitting read/write accesses.
In addition to the standard ATA interface, Trusted Computing Group (TCG) also
has its open standard named Opal/Opal2~\cite{opal} for SEDs.
What \sol needs is the fine-grained programmable write protection enforcement
(data secrecy not as a goal).

\section{Threat model and assumptions} \label{sec:threat}
\begin{enumerate}[leftmargin=1.5em]
\item We assume that malware/ransomware can acquire the highest software
privileges on a system (e.g., root/admin or even ring-0 on x86), through
any traditional mechanisms (often used by rootkits), including: known but
unpatched vulnerabilities, zero-day vulnerabilities, and social-engineering.
Root-level access allows malware to control devices (e.g., keyboard, network
interface), GUI, installation/removal of device drivers. 
\item Before deployment of \sol, the user system is not infected by any malware.
We primarily protect \emph{preexisting data} at the time of
malware/ransomware infection, and provide \emph{best-effort protection} thereafter for
later added/updated files until the malware/ransomware is detected (or a ransom is
demanded). 
	
\item We do not detect/stop the execution of malware, or \emph{identify} its actions. Instead, we protect integrity of
user data on a protected partition and ensure data accessibility. If the OS is
completely corrupted or inoperable, the user can install a new OS copy or boot
from another media (e.g., USB) \mbox{to access her data.} 

\item In the specific case of ransomware, we deal with the most common 
variants (i.e., cryptoviral extortion), and exclude those that simply lock access 
to system resources
without using encryption (non-encrypting ransomware~\cite{nonencrypt}) or
deletion, and those that threaten to publish information stolen from the user
(doxware or leakware~\cite{leakware}).

\item We assume all hardware (e.g., the CPU/chipset and the storage device), microcode/firmware and
other architecture-shipped modules (e.g., TXT's SINIT, see Section~\ref{sec:bg})
are properly implemented by the manufactures, and the user is motivated
to choose a system with no known flaws. An example of such a flaw is a series of
recently identified implementation bugs~\cite{sed-attacks} in SED firmware implementations that highly affect data secrecy (refer to Section~\ref{sec:sec} for details).

\item Attacks requiring physical access are excluded (e.g., no evil-maid
attacks). We only consider a computer system potentially infected by malware/ransomware
from the network or a removable drive.  

\item We assume that after infection, ransomware will act \emph{immediately};
i.e., it will find target user files, encrypt them, and then demand a ransom
without much delay (e.g., few minutes/hours, cf.~\cite{flashguard} vs.\ months). If the attacker waits, he risks of losing
control, e.g., through an OS/anti-malware update. 
With every
patched computer, the attacker loses money, and thus cannot remain hidden for
long. To accumulate file updates, the attacker may wait for
some time (i.e., long enough to collect sufficient content that the user may
care), before asking for the ransom. We term such attacks as \emph{persistent}
ransomware, and discuss them more in Section~\ref{sec:sec}, item (d).

\end{enumerate}

\section{Design}\label{sec:design}
We first define our design goals for \sol, then explore available/possible
choices with one or multiple TEEs placed at different locations of the storage
data flow and their implications; we illustrate our generalized design and its workflow by choosing readily available COTS options. We explain 
certain technical challenges/choices in Section~\ref{sec:impl}. 
The discussion will start with and be based on the stand-alone
mode of \sol. The network-based mode, which follows, just shifts 
the stand-alone complex of TXT/SVM, TPM and SED to a network location.
The terms ransomware and malware in our setting of unauthorized data alteration can be
interchangeably used. 

\subsection{Design goals}
We list our goals, and briefly sketch the key ideas to fulfill such goals 
in \sol.

\emph{a) Enforcement by device}.
Rootkit malware must not be able to modify or delete protected files. 
We place them in a write-protected mode (read-only)
all the time in the user OS. 
The write protection must not be bypassed or broken by 
rootkit and thus it must be
enforced by the storage device, where the protected partition
resides, instead of \emph{any software on the host}. Therefore, without the
appropriate authentication key (a high-entropy random value, e.g.,
256-bit long), the partition cannot be unlocked, even if the OS
is compromised (malware gains all software permissions).  

\emph{b) TEE-aided write protection}.
To allow updates, lifting the write protection is inevitable at a certain
point in time, during which protected files can be compromised by rootkit malware.
In \sol, write operations to the protected partition only 
occur inside a TEE. 
The authentication key is protected by, and bound to, this TEE
(inaccessible from outside).
All changes are treated as new versions (retaining historical versions)
and infrequent (batch) deletions are performed with user interaction or through policies. 

\emph{c) Minimal application interference}.
Applications (including the user OS) should operate as is.
As the original files are untouched by \sol and accessed the same way
by applications, normal application I/O is not hindered (even
for direct I/O as in disk utilities). File copies on the protected
partition are available as read-only, which should not concern regular applications.

\emph{d) Minimal user involvement}.
User experience should not be significantly affected.  A normal user experience
is preserved in \sol with the separation of the original and protected copies.
To reduce system unavailability for the stand-alone mode, the update/commit
process should be scheduled during \emph{idle} hours, and all updates to
the original copy are cached to be committed as a new version periodically
(e.g., every 8--12 hours). The user can also choose the network-based mode
as discussed in Section~\ref{sec:remote} to avoid unavailability.
The user is involved only when files must be deleted 
(sometimes including removal of old versions), and manually
triggering \sol (for immediate commitment of cached files, when the important
files are just edited/added). 

\subhead{Non-goals} \sol is designed to act more like a \emph{data vault} than a
traditional backup system; e.g., we commit user data a few times a day in batches,
instead of syncing updates instantly. 
Also, it adds another layer of protection to \emph{high-value user data}, complementing
existing backup systems. Namely,
OS/application binaries should not reside on the protected partitions;
regular user data may already be backed up (e.g., to certain cloud storage).
We provide robust data integrity against
advanced attacks at the expense of losing some data due to ransomware attacks
(e.g., user updates to a file during the commit period). Also, data
confidentiality is currently a non-goal (to facilitate unhindered operations of
common applications); i.e., the ransomware can read \emph{all} protected user
data, and read/modify the OS/unprotected partitions. However,
confidentiality and controlled read access can be easily supported; e.g.,
encrypting data under \sol-protected keys, and enabling password-based access
control for read operations on selected files. 

\subsection{Trusted file versioning} \label{sec:del}
We treat all write operations to the original copy (consolidated in one commit) as adding new
files to the protected partition (automatically approved, similar to S4~\cite{selfsecuring}), which poses
no threat to existing files, leaving only solicited file deletion (as opposed to version deletion) with user
intervention.  Any committed update to an existing file creates a new version,
instead of overwriting the current version (the latest one being under the
original file name) so that historical changes committed are all retained on the
protected partition.  
For space management, 
we leave it to the user to either clean up in the mini file browser we developed (see
Section~\ref{sec:scenarios} for details), or configure an auto-deletion policy
based on aging (e.g., after 1--2 years) or version-limiting (e.g., maximum 100 versions).  
The eventual choice is largely determined by the user's budget and needs.
Our simple versioning may not impose a significant
burden on the storage space, considering:
\begin{inparaenum}[a)]
\item We commit changes to the protected partition through scheduled invocation
of \sol; users can explicitly trigger the updater to commit important changes
immediately, which we believe would be infrequent. So the number of
versions that will be stored for a continuously updated file would still be
limited, e.g., 1--4 times a day. Auto-save in applications or file access-time
change do not trigger an update (it is only on the original copy).
\item Nowadays, disk storage is less costly and user computers are usually over-provisioned.
\end{inparaenum}
To improve storage utilization, specifically for large files, more space-efficient
versioning algorithms may be adopted (e.g., S4~\cite{selfsecuring}). Also, refer to the
file selection principles discussed below for optimal scenarios. 

User consent is mandatory when files (as opposed to versions) are to be deleted. 
File versions can optionally be deleted manually by the user.
We allow deletion operations in the mini file browser within the trusted environment,
where the user is asked to select explicitly which file(s) to delete. 
User consent is not needed in the case of auto-deletion of versions. 
Direct file
deletion in the protected partition outside the trusted environment will be ignored;
deletion of the original copy in the
unprotected partition will not be synchronized to the protected partition. 
We also hide old
versions from the user OS to help usability. When a new version is committed, we
rename the previous copy by appending its timestamp with the file name, and keep
the new version with the original name. 

\subhead{Automatic stale version deletion}
To relieve users from deleting unnecessary old versions of the same file, \sol
can be configured to automatically delete such versions 
after a certain time (aging) or number of versions (version-limiting). 
The retention duration should be long enough to hurt
ransomware's business model. For example, if an attacker needs to wait more than
a year to monetize his ransomware, it might become much less attractive than
now.  Defenders are likely to generate reliable detection mechanisms (e.g.,
signatures) within the wait period, and even be able to identify the attackers.
Calculation of the time duration (for aging) must be done appropriately, if there is no trusted
time source available within the TEE (e.g., TXT/SVM). As rootkits can change system time,
file creation/update time as available from the user OS file system cannot be
trusted. A simple solution could be to use digitally signed time value from an
NTP service,\footnote{See Section 6.2.2 at
\url{http://www.ntp.org/ntpfaq/NTP-s-config-adv.htm}. Alternatively,
time-stamping services, implemented by several CAs (following RFC 3161), can
also be used.} where the signature verification is done within the TEE. The signed
value can be obtained through the user OS, and must be sent for each file commit
session. The trusted updater must store the last accepted signed value along
with NTP verification keys, and check the new timestamp to detect replay (the
time value should always be increasing).

\subhead{File selection principles} 
Although technically we are not restricted in terms of file types, 
to optimize the user's budget and needs,
the user is recommended to follow a few principles:
\begin{itemize}
\item \emph{Targeted}. The user identifies what she 
cannot afford to lose (must pay ransom for). For example, an HD movie of 5GB
that can always be redownloaded should not be selected for protection. However,
we do consider such valid needs (e.g., of a movie editor), where corresponding
high-capacity disk space is assumed affordable.
\item \emph{Minimal}. The user should reduce redundancy from
the selected files. For instance, a mobile app developer may have 
(multiple copies of) the Android source tree (tens of GBs) which she compiles a couple of times
a day. The files changed by compilation are not necessarily those she  
cannot afford to reproduce. Therefore, she may just choose
source files reflecting her work.
\item \emph{Prioritized}. For instance, the update frequency for an important project
report (e.g., 1--3MB) should be set higher in the policy than the user's favorite
songs (e.g., MP3s, 300MB).
\end{itemize}

\subsection{TEE placements} \label{sec:multitee}
Based on our design goals, \sol can be hypothetically constructed around
the following entities: 
\begin{enumerate}
\item \emph{Host}. The processor-centric system where software (including user
applications) runs and where data is generated/consumed. Benign and malicious
operations cannot be naturally distinguished. 
\item \emph{Disk}. The storage device where legitimate user data (from user applications)
is supposed to be securely stored. No additional mechanism other than 
physical safety is assumed. 
\item \emph{TEE}. There might be multiple TEEs. They can be equipped on either
the Host or the Disk, or both. The TEE has the properties and purposes discussed in Section~\ref{sec:bg}.
In light of the Disk's constraints, we further model all TEEs for three types of functionalities: 
\begin{inparaenum}[I)]
\item Policies. This applies to general-purpose TEEs (as with that of the Host), where the user/vendor can
provision arbitrary storage rules, e.g., write protection.
\item Authentication. We consider the most constrained application-specific TEEs on the Disk
to be able to authenticate commands from the Host with a configured secret, e.g., a TLS certificate and/or a password.
\item Secure protocol. A secure communication protocol between the Host and the Disk may be supported (in addition to authentication), which ensures integrity and confidentiality, e.g., the untrusted OS cannot learn what is being sent and if it tries to manipulate the data it will be detected.
\end{inparaenum}
\item \emph{Interface}. This is how the Disk is connected to the Host physically, as exemplified by
SATA, USB and Ethernet. We consider both secure and insecure interfaces.\footnote{A secure interface ensures data integrity and confidentiality (but not necessarily authenticity) in the discussion hereinafter.}
In a threat model with no physical access, direct links with endpoints inside the TEE 
can be treated as secure. Otherwise, for example, Ethernet (WAN, where other nodes 
are involved) or SATA relying on untrusted kernel drivers is susceptible to interception.
A secure interface cannot provide message authenticity if it is not \emph{dedicated} to the TEE.
For instance, outside the TEE (for concurrent TEEs) or when the TEE is not active (for exclusive
TEEs), untrusted software on the Host can still send anything to the Disk, unless there is a way to 
block access (e.g., with the TrustZone protection controller [10]) to the Interface from untrusted 
software at all times.
\end{enumerate}
The storage data flow goes between the Host and the Disk through the Interface.
The TEE's location can shift along the data flow across different entities.
In addition to the autonomous enforcement of predefined policies (e.g., append-only),
certain \emph{administrative access} to the Disk is also needed for space management,
i.e., there must be a convenient way for the user to delete files or change the configuration.
In the following, we consider a few possible constructions with different 
TEE placements. We use the subscript to indicate where the TEE resides,
e.g., \emph{TEE\textsubscript{Disk}} means a TEE is on the Disk.

\begin{figure}
\centering
	\includegraphics[width=0.47\textwidth]{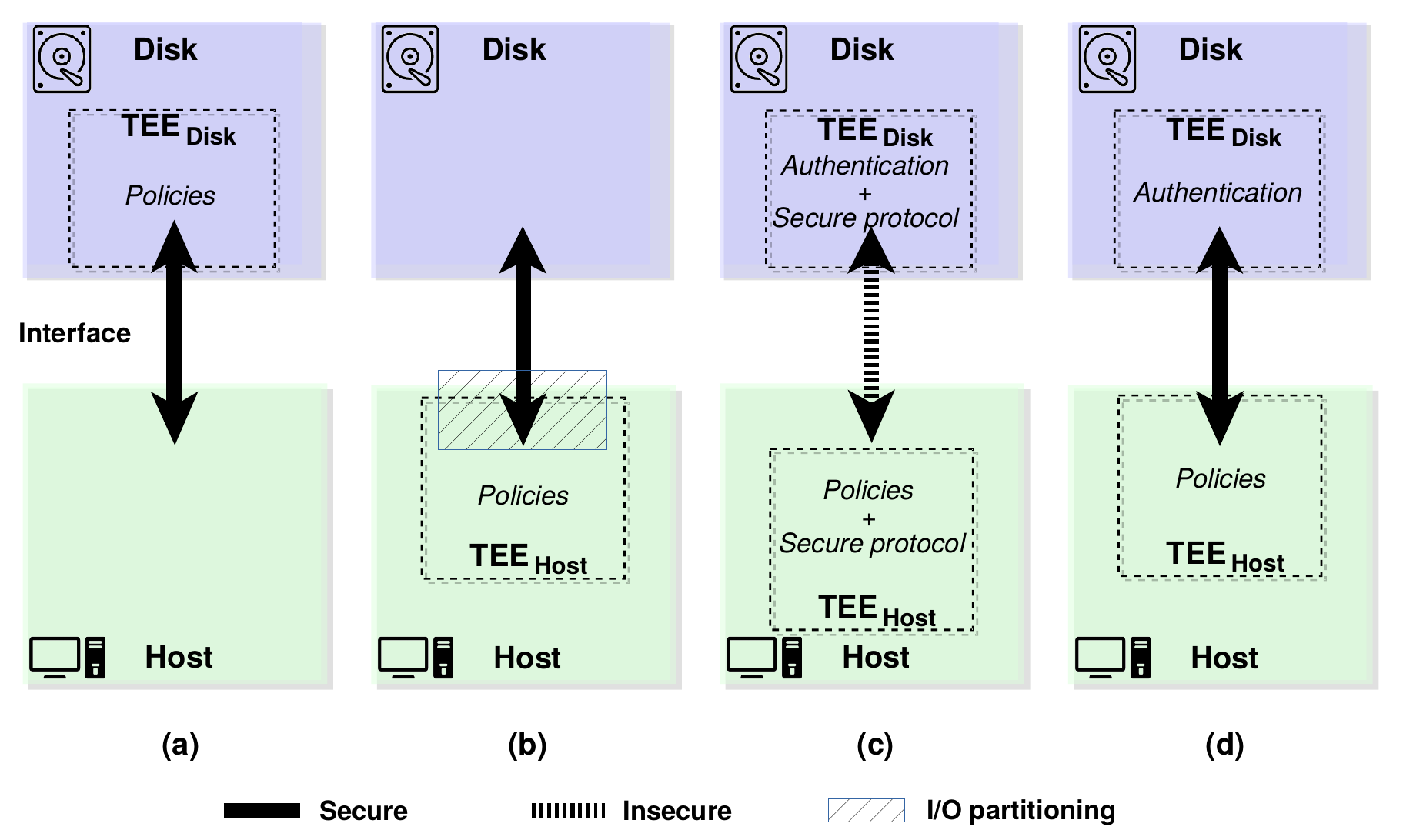}
	\captionof{figure}{Example constructions with varying TEE placements. An 
\emph{Interface} arrow not terminating inside \emph{TEE\textsubscript{Host}} means the TEE is unprivileged and
relies on other software (e.g., the OS) to perform I/O. Privileged TEEs (with the arrow inside) 
only own \emph{Interface} when active, hence the need for certain I/O partitioning (e.g., TZPC) to
block it to other software when the TEE is not active, making \emph{Interface} dedicated}
	\vspace{-20pt}
	\label{fig:MultiTEE}
\end{figure}

\subhead{TEE\textsubscript{Disk} alone}
If \emph{TEE\textsubscript{Disk}} enforces the append-only and
history-preserving logic on its own, \emph{TEE\textsubscript{Host}} is not necessary.
Figure~\ref{fig:MultiTEE}-a depicts this construction.
However, it lacks a trusted user interface for the administrative access. 
This might be solved by adding 1) a physical switch, to allow occasional 
full access, or 2) a secondary dedicated interface, assuming there exists
a trusted host for performing administrative tasks.
Therefore, we do not consider this construction for our prototype. \looseness=-1

\subhead{TEE\textsubscript{Host} + secure Interface}
In contrast to \emph{TEE\textsubscript{Disk} alone}, if there is only
\emph{TEE\textsubscript{Host}} enforcing the logic, regardless of whether
the Interface is secure or not, our goals still cannot be achieved.
For insecure Interface, malicious enitity can directly intercept traffic and corrupt data.
For secure Interface, which might seem sufficient for channel integrity and secrecy, 
untrusted software on the Host can also do harm at the data origin.
When \emph{TEE\textsubscript{Host}} is not in control, other privileged code
can access the secure Interface and modify/delete the stored data.
The root cause is that the Disk has no TEE to authenticate \emph{TEE\textsubscript{Host}}.

\subhead{TEE\textsubscript{Host} + dedicated secure Interface}
On top of the construction above, if \emph{TEE\textsubscript{Host}} comes with an I/O partitioning mechanism
(e.g., the TrustZone protection controller),  the secure interface can be configured
in a way that when an exclusive \emph{TEE\textsubscript{Host}} is not active or outside a concurrent \emph{TEE\textsubscript{Host}}, no software  can access it (Figure~\ref{fig:MultiTEE}-b).
As an advantage, this allows any storage device to be used as the Disk. 
Also, with \emph{TEE\textsubscript{Host}}, the administrative access can be provided to the user. This construction may become  feasible in the future when ARM-based desktop/laptop computers can efficiently run modern desktop OSes~\cite{arm-server}. 

\subhead{TEE\textsubscript{Host} + TEE\textsubscript{Disk} + insecure Interface}
As \emph{TEE\textsubscript{Disk}} is usually not general-purpose (i.e., supporting arbitrary code to be provisioned),
we need \emph{TEE\textsubscript{Host}} to be used in conjunction to achieve our purpose.
A minimum support from such \emph{TEE\textsubscript{Disk}} 
is the use of certain secret for encryption or access control (e.g., SED and Kinetic storage), which can
serve as the basis for write protection (read-only). If \emph{TEE\textsubscript{Disk}}
supports a secure communication protocol, e.g., TLS, a secure Interface is unnecessary. 
In this case, as long as the shared secret (or its public-key equivalent)
is properly handled in \emph{TEE\textsubscript{Host}}, a secure channel is formed, 
equivalent to a secure Interface. \emph{TEE\textsubscript{Host}} enforces all the logic.
Note that unprivileged TEEs (see Section~\ref{sec:bg}) can be used here,
since the requirement for a secure Interface has been relaxed.
See Figure~\ref{fig:MultiTEE}-c for an illustration. This corresponds to the construction of Pesos~\cite{pesos},
as detailed in Section~\ref{sec:rel}.
This construction is also missing a trusted UI for administrative access 
(forged input can be entered in \emph{TEE\textsubscript{Host}} thus
deleting unintended files), as unprivileged TEE only interacts with the user
through the untrusted OS.

\subhead{TEE\textsubscript{Host} + TEE\textsubscript{Disk} + secure Interface}
Peripherals directly connected to a PC with exclusive/privileged \emph{TEE\textsubscript{Host}}
is considered to have secure Interface. In this case,
\emph{TEE\textsubscript{Disk}} that does not support a secure communication protocol 
is applicable (e.g., Opal SED). \emph{TEE\textsubscript{Disk}} authenticates 
\emph{TEE\textsubscript{Host}} in cleartext; this could be problematic if attackers gain physical access or \emph{TEE\textsubscript{Host}} is non-exclusive/unprivileged. Privileged \emph{TEE\textsubscript{Host}}
provides a trusted UI for administrative access (forged UI outside \emph{TEE\textsubscript{Host}}
cannot be authenticated by \emph{TEE\textsubscript{Disk}} without the secret).
We choose this construction to implement \sol; see Figure~\ref{fig:MultiTEE}-d. 

\subsection{Design choices} \label{sec:choices}
According to the discussion above, we expand on further considerations and 
requirements on the chosen construction of \emph{TEE\textsubscript{Host}} + \emph{TEE\textsubscript{Disk}} + secure \emph{Interface}.
Note that the design choices are only based on available COTS devices for this prototype.
The high-level design is generalizable to other \emph{TEE\textsubscript{Host}} and \emph{TEE\textsubscript{Disk}} (as shown 
in Section~\ref{sec:multitee}).

\subhead{Host-side TEE}
To satisfy secure Interface, \emph{TEE\textsubscript{Host}} must be privileged (Figure~\ref{fig:MultiTEE}-d);
otherwise, the OS may manipulate the I/O traffic. Also, our observation shows that
if a TEE is privileged, it is usually exclusive (i.e., not allowing other software to run in parallel)
unless there is hardware I/O partitioning as with ARM TZPC, to avoid resource access conflict
or contention. Actually, as an advantage with privileged TEEs, 
no other applications (including the OS) can have the chance to
even observe what is running inside, as anything else is discarded/suspended, avoiding
software side-channel attacks.

We use Intel TXT/AMD SVM as the host-side TEE
for \sol. Therefore, our discussion hereinafter will refer to \emph{TEE\textsubscript{Host}} as TXT/SVM.
The more recent Intel SGX does not run privileged code for I/O access.\footnote{
SGX has also faced several side-channel attacks due to its concurrent nature, e.g., Foreshadow~\cite{foreshadow}, Branch Shadowing~\cite{branchshadow}, and cache attacks~\cite{sgxcache}.} 
\looseness=-1

\subhead{Programmable write protection}
As the ideal \emph{TEE\textsubscript{Disk}} (a general-purpose on-disk TEE, Figure~\ref{fig:MultiTEE}-a) 
is unavailable (as of writing),
and it also lacks a proper trusted UI on the Host for administrative access,
we resort to storage devices with an application-specific TEE.
\emph{TEE\textsubscript{Disk}} needs to be able to communicate with \emph{TEE\textsubscript{Host}} for 
the write protection (programmability). We expose write access to the protected 
partition only inside \emph{TEE\textsubscript{Host}}. 
Some off-the-shelf secure USB drives offer write protection~\cite{usbdrive}.
However, it is either in the form of a physical switch/button to be pressed by
the user, or a key pad on the USB device itself, where a password can be typed 
(like a closed TEE inaccessible from outside).

The self-encrypting drive (SED, see Section~\ref{sec:bg}) satisfies
programmability, with one or more secrets for authentication or media encryption.  
Also, SED has the advantage of supporting
fine-grained protection ranges with separate read/write permissions, which is
important as we constantly allow read access, and deny write access from the user
OS. Fine protection granularity also allows the protected partition to coexist
with the unprotected OS and other files in the same drive, instead of requiring
a dedicated disk. The only disadvantage, as far as \sol is concerned, is that
SED does not support secure communication protocols thus requiring a secure Interface.
The legacy ATA Security password can also be considered 
device-enforced write protection (without media encryption). However, it is 
a non-solution for \sol, because only one-way locked-to-unlocked transition is 
allowed (SEC4:SEC5~\cite[p.~13]{ata-security}), i.e., relocking requires power reset, whereas \sol needs 
the ability to switch back and forth. \looseness=-1

\subhead{Minimal TCB}
Although a full-fledged OS in TXT/SVM (e.g., tboot with Ubuntu) can be used to
perform trusted operations, it is preferable to keep a minimal trusted computing
base (TCB), for both auditability (e.g., avoiding numerous complex components)
and maintainability (e.g., avoiding measuring large and varying files). Moreover, it is
technically more involving, because the trusted operations occur in the midst
of an active user OS execution (considering the time/effort needed to save and
restore various states for both OSes).  
Therefore, we develop our own logic as a small-footprint, native
program in TXT/SVM with no external dependencies.

\subhead{Separation of the protected partition from the original}
Technically, we can write file updates immediately on the protected partition.
However, unsolicited/frequent write attempts, such as updates from the automatic
save feature in text editors (i.e., not initiated by the user clicking on the
``Save'' button), will create too many versions on the protected partition and
make the system unusable due to frequent switch between regular and
trusted environments; note that, TXT/SVM is exclusive, and writing file updates may
also take noticeable time.  Therefore, 
we leave user-selected files where they are, and make
a copy onto the protected partition on the SED. All subsequent updates happen
to the original files without write protection.
The user can then decide when to commit
changes to the protected partition (no versioning on the original partition),
manually, or automatically at certain intervals (e.g., every 8-12 hours).  \looseness=-1

\subhead{File-system in TXT/SVM}
For protected write operations, we cannot simply pass the raw sector information
(sector number, offset, number of bytes and the buffer) to TXT/SVM as we perform
file-based operations, and the user also must select files (not
sectors) for deletion. Therefore, the TXT/SVM program must be equipped with a file
system.

\subhead{Data mobility and recovery}
The SED can also contain an unprotected partition where the OS resides, 
because of the fine granularity of protection ranges, while sometimes users may treat it as a
stand-alone data drive. 
In either case, when data recovery is needed (e.g., the OS is corrupted or compromised), 
the user can simply 
reboot from different media on the same machine or mount the SED on a different
machine. The data will be readily accessible as read-only, hence aiding data mobility, 
thanks to the separation of read and write accesses.
In case the user needs to update the files, a rescue USB, where all intact \sol binaries
are stored as well as a portable OS can be used to boot the same computer (where \sol was provisioned). After booting with 
the rescue drive, the
user can invoke the same updater in TXT/SVM for regular file access or
deprovisioning (to remove the write protection).

\subsection{System components and workflow}   \label{sec:sys}
\begin{figure}
\centering
	\includegraphics[width=0.47\textwidth]{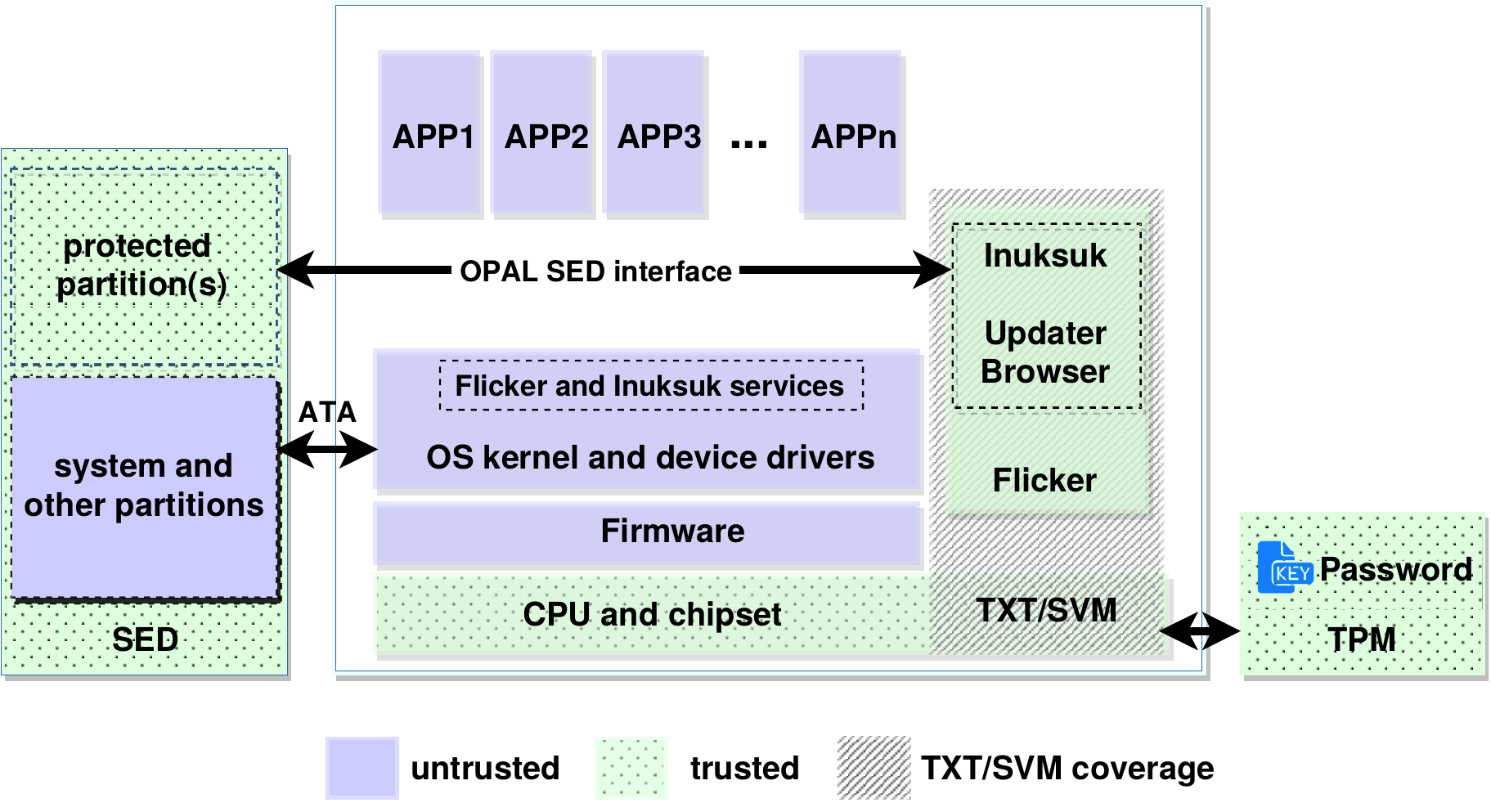}
	\captionof{figure}{System overview}
	\vspace{-20pt}
	\label{fig:Overview}
\end{figure}

Refer to Figure~\ref{fig:Overview} for an overview of our design.  The system
consists of the following components at a higher level (further technical details
are discussed in Section~\ref{sec:impl}): 
\begin{itemize}[leftmargin=1.1em]
\item \emph{Trusted updater.} This is the core component of \sol, and runs
inside TXT/SVM. It is responsible for copying files from the original partition to
the protected partition (in SED write access mode) as new versions, file listing
(in a mini file browser), and showing file meta data to the user. Files in the
original partition are selected to commit based on their last-modified timestamp.
\item \emph{TPM.} In conjunction with TXT/SVM, TPM makes sure that the secret (the
SED password) is securely stored in its NVRAM storage, and can be
\emph{unsealed} only if the unmodified trusted updater is executed (as
\emph{measured} in TPM's platform configuration registers). 
\item \emph{Secure drive.} An SED drive hosts the protected partition.
Without the high-entropy key/password, its protection (i.e., write protection 
in our case) cannot be bypassed. Note that even with physical access to
the drive, reinitializing the drive with the PSID (physical secure ID) printed on it will have
all data lost. \looseness=-1
\item \emph{OS drivers.} A few OS-dependent modules are needed to bridge the
user, OS and the trusted updater, such as preparing the TXT/SVM environment. 
These modules do not have to be trusted after
initial deployment, as the worst case is a DoS attack; see
also Section~\ref{sec:sec}, item~(b).
\end{itemize}

\subhead{Update policies}
The update policies reflect user preferences and control how the trusted updater behaves
(as primitive text files in the prototype for now), e.g., the age threshold, 
version limit, maximum file size, scheduled interval, etc. Moreover, the user can
optionally configure certain checks to be enforced in TEE for the purpose of anomaly
detection, e.g., ransomware maliciously triggered 100 versions of a protected file.
At each run, a report is generated securely in TEE, e.g., showing a list of updated files. \looseness=-1

The policies are sealed on the protected partition (can only be unsealed inside TXT/SVM).
A plaintext copy is left on the unprotected partition for certain operations (e.g.,
scheduled updates based on an interval). Once in the trusted updater, the unsealed copy of policy
can be used to verify the action's correctness (e.g., manipulated interval outside).

\subhead{Workflow}
The generalized workflow of \sol is as follows:
\begin{inparaenum}[(a)]
\item At deployment time, a high-entropy secret is generated as the SED password
and sealed into TPM (can only be unsealed in the genuine trusted updater). 
\item The protected partition is created with the SED write protection. 
The user also
selects the files to be protected, which are then copied to the protected
partition in the first invocation of the updater. After the first-time
copying, the user still interacts directly with her files on the original partition.
\item In everyday use, the protected partition is never touched (except for read-only access). 
As with certain cloud storage services, we use an icon on the original files 
to indicate which ones are under the protection of \sol.
\item If the user adds or updates files on the original partition and is ready to commit her changes, she triggers the trusted updater, and without involving
her to verify, changes are committed as new files/versions on the protected
partition. The updater is triggered either
manually, or automatically, e.g., via scheduled tasks, when the updating-application is closed,
or when the system is restarting or shutting down.
\item When the user wants to delete files or old file-versions, she can manually
trigger the updater to open a mini file browser, and make the selections. 
\end{inparaenum} \looseness=-1

\subsection{Network-based data vault} \label{sec:remote}
The functionality of \sol does not rely on any third parties (except
the device manufacturer), as the trust is anchored in hardware/firmware and 
all its components are local. Although our explanation of \sol is based 
on its stand-alone mode, there is no fundamental barrier in 
the design for it to be deployed as a remote/networked data vault.  
To provide users with a centralized network-based mode, as well as extending
for enterprise and cloud storage services,
we briefly explore a variant of \sol where the key components, i.e., TXT/SVM CPU,
TPM and SED, are shifted to a network location, forming a remote service. 
Users' data will remain protected at a central, \sol-backed storage service, 
and users can keep using any device of their choice (i.e., with or without TEE, 
mobile or desktop, at home or in an enterprise). 
We believe that this variant can be used to protect security-sensitive/user 
files stored in cloud storage services like Dropbox and OneDrive, or 
enterprise storage services. Although such services are possibly backed 
by robust backup measures and strict security policies/tools, if infected, 
consequences can be high. 

The construction goes as follows: any desktop/laptop/mobile device
serves as the \emph{front-end} directly used by the user. Through an account,
the front-end is connected to a \emph{storage back-end}, which plays the role
of the ``original partition'' in our stand-alone setup, caching file updates.
Eventually, an \emph{\sol-equipped backup server}, which has the TXT/SVM-capable
CPU and chipset, as well as the SED (or more likely, an SED array), is connected 
with the storage back-end. The \sol-server will periodically copy new/updated files from the storage server, and become unavailable during this period, which should not affect functionality, assuming the \sol-server is not used for other purposes. The storage back-end and user devices remain available all the time.
 
Once deployed correctly, without the high-entropy key sealed in TPM, no remote attacker can turn off the write protection
and update/delete the protected files. Our threat model now assumes that the remote attacker can infect the storage and \sol servers, in addition to user devices. As before, only the uncommitted files remain vulnerable, and after written to the \sol protected storage, user files become safe against any data modification attacks. 
Content on the \sol-equipped server can be maintained by enterprise IT administrators (e.g., for deleting old versions).
The whole process is transparent to end-users/employees, and the files that need ransomware protection can be identified by enterprise policies.
In the case of home users, the storage back-end and the \sol-equipped server can physically reside on the same NAS (Network Attached Storage) device,
assuming certain disruptions are tolerable.

\section{Implementation and evaluation} \label{sec:impl}
We have implemented \sol for Windows 7 non-PAE, Windows 10 64-bit and Ubuntu 12.04. 
In this section, we discuss important technical considerations
and choices, as well as performance evaluation/issues during the prototype implementation.
Our techniques can also be useful for other OS-less I/O intensive TXT/SVM applications.
We perform our development and testing on an Intel Core i7-2600 @3.40GHz
(and AMD FX-8350 8-Core @4.0 GHz), 16GB RAM, and Seagate ST500LT025
SED disk. The performance numbers are from the Intel machine. Also a few
other computers are used for debugging and cross-validation. As an estimate
of the TCB size, in addition to Flicker's codebase (2012-06-18, v0.7), \sol adds
5190 LoC (in the PAL, which runs in TXT/SVM, OS drivers excluded). 
Among this, fat\_io\_lib contributed 1852 (for FAT32), followed by TCG Opal (1507) and USB (1467).
We used the tool LocMetrics~\cite{loc} to count the lines with some trivial manual effort.
There are also additional 
engineering-oriented technical details omitted here for brevity, such as resource handsoff 
with the OS for secure user interface (keyboard and display).

\subsection{Porting and using Flicker for TXT/SVM sessions}
Since \sol's secure file operations occur alongside the user OS, a mechanism is
required for jumping back and forth between the trusted updater and the user OS.
It can be implemented as a device driver (in the user OS) dealing
with parameters, saving the current OS state, processing TXT/SVM logic, and
restoring the saved OS state when returned from the trusted updater. Several
such operations are already handled in Flicker~\cite{flicker} (also refer to Section~\ref{sec:bg}), 
which we use as the
base of our prototype.  

Flicker supports only 32-bit non-PAE Linux and Windows 7 OSes (no update 
since 2012), which is a limitation for our prototype with modern 64-bit systems.
We thus port Flicker to the latest Windows 10 64-bit, which may benefit
the community for other in-OS secure processing with device I/O, or when 
the exclusive processing environment of TXT/SVM is desired (as opposed to 
other TEEs such as SGX and TrustZone). Similar to the 32-bit Flicker, our adaptation, 
especially for context saving/restoring, is also mostly based on heuristics, as
the available Intel/AMD documentation does not specify what is preserved or
affected by TXT/SVM. Here, we omit the technical challenges for this porting,
for example, the 4-level unity page table covering both physical and virtual, 
32-bit and 64-bit addresses; 64-bit context saving/restoring; the discontinued 
inline assembly, etc. Note that this process is far more complex than 
adapting user-space applications or kernel drivers/modules for 64-bit.

\subsection{Opal access to SED inside TXT}
All software outside TXT/SVM, including the OS and all its device drivers, is
untrusted in \sol. However, inside the TXT/SVM environment, there is no run-time
device support, i.e., devices including any SED drive cannot be accessed by
default.  Therefore, we must implement standalone (and preferably lightweight,
to limit the TCB size) custom driver for accessing SED devices inside TXT/SVM.  Various
SED protocols rely eventually on the SATA interface (ATA Command
Set~\cite{ata}), with two options to choose from: 
\begin{inparaenum}[1)]
\item ATA Security password~\cite{ata-security} (e.g., prompted in BIOS at bootup). 
In this option, SEDs only differ
with regular hard drives in that user data is encrypted on the media.
\item Dedicated security protocols (e.g., Seagate
DriveTrust~\cite{drivetrust}, IEEE 1667~\cite{ieee1667} and Microsoft eDrive
(all based on TCG Opal/Opal2~\cite{tcg-opal}). They implement support for multiple roles/users
corresponding to multiple ranges, with separate passwords for write/read access.
\end{inparaenum}

Granularity in both protection ranges, and separate read/write permissions
is important in our design. The same drive can
host both protected and unprotected partitions (which cannot be achieved in
Option 1).
Thus Option 2 is more suited for our needs, and we choose to use TCG Opal to
communicate with SED, as it is an open standard and widely supported by
most devices. Actually, with TCG Opal we can define multiple ranges (corresponding
to partitions) protected by different secrets. Then these secrets can be sealed
with different programs (e.g., the updater is one) thus reducing the risk of exposing
all protected data altogether as in a single point of failure. We leave this as future work.
Note that Opal is merely the payload security protocol (SFSC) of the carrying interface,
be it SATA or NVMe; NVMe supports two variants of TCG Opal, Opalite and Pyrite~\cite{nvme}.
We anticipate that supporting NVMe (for higher performance) may require only trivial changes. \looseness=-1

A few open-sourced tools can
manipulate SED devices with OS support (in addition to proprietary tools for vendor-specific
protocols); we have tested msed~\cite{msed} (now merged into DTA
sedutil~\cite{sedutil}) and topaz-alpha~\cite{topaz-alpha}. They mainly rely on
the I/O support from the OS, e.g., SCSI Generic I/O, in the ATA passthrough
mode.  However, our TXT/SVM PAL is OS-less with no
run-time support. We decide to port functions from
topaz-alpha~\cite{topaz-alpha} as needed.  The porting process faces
several engineering challenges, which we omit here for brevity.

\subsection{File system efficiency}
We handle updates to the
protected partition at file-level instead of raw sectors (see Section~\ref{sec:design}). 
This requires at least
basic file system functionalities implemented within TXT/SVM. 
To avoid rework, we tested several libraries, including fat\_io\_lib~\cite{fatiolib},
ThinFAT32~\cite{thinfat32}, fedit~\cite{fedit}, efsl~\cite{efsl}, etc.
FAT32 projects that are tightly coupled with external dependencies are
excluded (e.g., the mainstream FAT32 support with Linux VFS \texttt{inodes}).
But none satisfies both the two necessary features:
\begin{inparaenum}
\item \emph{Buffering support.} Usually, FAT32 access is sector-wise, while DMA
requests need to handle as many sectors as possible to reduce per-request overhead.
PIO access is not affected but it is by nature slow.
Note that hardcoded pre-fetching for reads is an overkill (reading
data never needed), and hardcoded write buffer will hang (waiting for enough
number of sectors).
\item \emph{Multi-cluster support for space allocation.} At the file creation
time, and when a file grows in size, FAT32 must traverse all clusters to find
free clusters to be appended to the cluster chain of the file.  
Interestingly, with all FAT32 projects we tested, only one
cluster is allowed to be added (we do not see any performance problem for
allowing multiple).  Therefore, for a 50MB file taking 6400 clusters (8KB
cluster-size) and the partition having 131072 free clusters (1GB), it takes more
than 800 million iterations. \looseness=-1
\end{inparaenum}

We choose fat\_io\_lib for adaptation, because of its good buffering
performance. To add multi-cluster support, for each iteration, we start with the cluster where we left off,
instead of the first cluster of the partition. 
We emphasize that \sol is not
dependent on any specific file system, and thus FAT32 can be replaced with a
more efficient one.

\subsection{Discussion on DMA inside TXT} \label{sec:dma}
The necessity for Direct Memory Access (DMA) is ubiquitous, even for a light-weight
program like the \sol updater. For instance, USB keyboards are the defacto norm.
Unlike other simpler protocols, the controller (e.g., EHCI~\cite{ehci})
requires several host-allocated buffers in the main memory (DMA chunks) for
basic communication with the host (e.g., the periodic frame list). The
controller accesses the buffers without the CPU's intervention, hence, direct
memory access. Also, for data transfer to/from the hard drive (e.g., SED),
The theoretical speed of ATA PIO modes~\cite{pioperf} is very low. Taking into
account the file system overhead, it needs 3--4 minutes to write a 100MB file,
which is unacceptable from the user-experience perspective. In modern systems,
DMA is usually enabled for disk access. \looseness=-1

However, the fundamental protection of TXT (like all other
TEEs) must prevent autonomous access from peripherals for the protected regions.
The MLE (Measured Launch Environment, specific to Intel TXT) memory is 
included in either the DMA Protected Range (DPR) or 
Protected Memory Regions (PMRs), which is mandatory (cf.~\cite{mle}).
Consequently, since we cannot (and do not want to) exclude the MLE from DMA
protection, we have to allocate the DMA chunks outside. In our specific case where 
physical attacks are excluded and no other code is running in parallel, exposing DMA 
regions outside the MLE does not pose a threat (more in  
Section~\ref{sec:sec}, item (e)).

In this way, we implement our custom DMA support for both the 
USB keyboard and ATA DMA controller for data transfer. We also support PS/2 keyboard. 
Usually, DMA relies on interrupts, i.e.,
when the transfer is done, the interrupt handler will be notified to proceed to
the next request (e.g., to maximize CPU time utilization in a multitasking
environment).  In our case, Flicker is not supposed to work with an
interrupt-enabled workload (technically possible with some complex adaptation),
and we merely need the performance boost through DMA, i.e., no
multi-tasking and thus, requiring no interrupt support. Therefore, our custom
DMA support works with polling. With DMA
enabled, file transfer in the \sol updater is 50--100 times faster than 
using just PIO (see Section~\ref{sec:scenarios}). \looseness=-1

\subsection{Measuring disruptions}    \label{sec:scenarios}
\begin{figure}
\centering
	\includegraphics[width=0.4\textwidth]{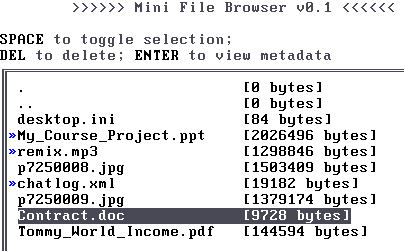}
	\caption{A screenshot of the mini file browser inside the trusted updater.
Selected files are designated with ``\guillemotright''; group selection can be specified by the first and last files.}
	\vspace{-20pt}
	\label{fig:minibrowser}
\end{figure}

\sol does not introduce run-time performance overhead for user applications.
However, when updates are being committed to the protected partition (file
copying), the computer will be unavailable for regular tasks, due to TXT/SVM's
exclusiveness. Such disruption is determined by both the
file I/O performance and various factors, e.g., the file count/size.
It mainly comes from data transfer for new versions, i.e., created/modified files.
Deletion involves only flagging the files as deleted in the file system, and thus is quick. 
We have developed a light-weight file browser inside the trusted updater that
allows the user to choose multiple files for deletion; see
Fig.~\ref{fig:minibrowser}. With more engineering effort, graphical interface
can also be created. There is no technical limitation for creating custom UI
within TXT/SVM.
Also recall that by configuring \sol in
the network-based mode, such disruption can be mitigated.
The discussion here only focuses on the stand-alone mode. \looseness=-1

\subhead{File I/O evaluation}    
The file transfer speed determines the duration of disruptions, and
affects user experience. However, we argue that the way we implemented DMA and
our choice of the FAT32 library (as well as our adaptation to it) are confined
by the engineering effort and time. Therefore, the numbers we show here should
be just the lower bounds.

As micro-benchmarking, we executed 10 measurements on the files we selected with fixed sizes; see
Table~\ref{tab:numbers}. They are all one-way access, write-new/write-existing/read respectively.
50MB represents common media files and 500KB represents miscellaneous files of trivial sizes.  Note that
without our adaptation to enable multi-cluster allocation support, the creation
of a 50MB file can be done only between 0.5--1MB/s, while overwriting of an
existing file of the same size runs at about 40MB/s. To have a rough comparison
with regular file transfer performance, we also list the corresponding speed in Windows aside
(a 10-time average of reads and writes). The overhead percentage is calculated with
the lowest speed of \sol divided by the corresponding Windows value.

To demonstrate \sol's performance in a realistic usage scenario, we invoke the trusted
updater to copy 50 random photos (JPG files, size ranging from 1009KB to 2416KB,
totaling 85.6MB) 
from the original partition to the protected partition. We measured the duration
for 10 times, and the performance seems reasonable (mean: 23.3853 seconds), and relatively 
stable (standard deviation: 0.58989). This is a combination of read, write, file
opening/closing, accumulating space fragments, etc. We also evaluated only the transition
time between the OS and trusted updater.
It varies between 2--4 seconds, including screen mode switching.

If we take into account any extra processing during file transfer, the time needed
may also be affected. The basic versioning \sol uses is not \emph{incremental}, i.e., 
the whole of the source file on the original partition is copied over to the
protected partition as a new version. We may consider some open-source
version control systems like SVN/Git (or even the simple \texttt{diff} command)
for incremental versions to save disk space. However, in that case,
each time new files/updates are committed, the updater must scan the whole
of both files for differences and then perform the transfer. Moreover, deletion
is supposed to be very quick with non-incremental versioning (just flagging
the file); with diff-like versioning, for each file the updater has to reassemble from
all previous versions to form the latest one to be kept. The overhead could be
significant in our setting (considering batch-deletion of versions). Also, for
common file types such as images, videos, and rich documents (e.g., PDF, Word),
incremental versioning may not save much disk space.

\begin{table}
	\centering
	\resizebox{0.47\textwidth}{!}{%
		\begin{tabular}[c]{cc|d{2.2}|d{2.2}|d{2.2}|d{2.2}|d{2.2}}
& & \multicolumn{1}{c|}{\text{Write/Existing}} &
 \multicolumn{1}{c|}{\text{Write/New}} & \multicolumn{1}{c|}{\text{Read}} & \multicolumn{1}{c|}{\text{OS}} & \multicolumn{1}{c}{\text{Penalty}} \\ \hline
\multirow{2}{*}{50MB file} & \multicolumn{1}{|l|}{Mean}
& 43.93 & 41.69 & 32.17 & 68.77 & \multicolumn{1}{c}{\text{53\%}} \\ 
& \multicolumn{1}{|l|}{SD} 
& 3.40 & 0.31 & 0.09 & & \\ \hline
& \multicolumn{1}{|l|}{Mean} 
& 26.46 & 8.09 & 16.67 & 14.29 & \multicolumn{1}{c}{\text{43\%}} \\ 
\multirow{-2}{*}{500KB file} &
\multicolumn{1}{|l|}{SD} 
& 1.18 & 0.43 & 5.26 & & \\ \hline
\end{tabular}}
\caption{File transfer performance (MB/s) in the trusted updater from 10 measurements. For small files (e.g., 500KB), other overhead predominates the transfer time.}
\label{tab:numbers}
\vspace{-20pt}
\end{table}

\section{Security analysis} \label{sec:sec}
In this section, we list various potential attack vectors, and discuss how they
are addressed, or why they do not pose a threat (see
Section~\ref{sec:threat} for our assumptions).

Since we shift the defense from detection/recovery to data loss prevention,
we avoid common attacks such as
whether new malware/ransomware can evade detection, whether it does privilege escalation, and how the 
encryption keys are generated. There are two basic questions in 
evaluating \sol's effectiveness:
\begin{inparaenum}[1)]
\item Outside the trusted environment, can malware/ransomware update files on the
protected partition? No, without the high-entropy key sealed in TPM, software
on the host system cannot break the write protection enforced by SED.
\item Inside the trusted environment (updater), can the malware/ransomware trick the
user or the updater to write arbitrary content? The updater does
not synchronize any file deletion from the original partition but only adds
files from it. With the updater's integrity ensured by TXT/SVM, user I/O cannot be
influenced by any external software. 
\end{inparaenum}

\begin{enumerate}[label={\bf(\alph*)},ref={(\alph*)},itemsep=2pt,leftmargin=0cm,itemindent=.5cm,labelwidth=\itemindent,labelsep=0cm,align=left,listparindent=\parindent,
parsep=0pt]

\item
\bfdot{Forged user interface}
Due to human users' inability to authenticate machines (cf.\
Stark~\cite{stark}), rootkit malware may mimic the
appearance of the intended application, where the user may leak
secrets. However, the adversary will not benefit from it, as there is no UI in
\sol for prompting for the SED unlock secret (in fact, the unlock secret is
unknown to users).  Also, for manual deletion, there is no way to specify which
files to delete from outside the trusted updater (files are selected in TXT/SVM
right before they are deleted). In the end, without the genuine updater in TXT/SVM,
the adversary cannot manipulate any file on the protected partition.

In addition to leaking secrets, forged UI can also help malware hide termination
(skipping execution of \sol), making the user believe the update has occurred (actually it was a fake one).
To discern, the user can configure a per-deployment secret such as an
\emph{avatar} or \emph{short phrase} to be shown in the trusted updater (mini file browser).
Because of the exclusiveness and I/O isolation of TXT/SVM, rootkit malware
will not learn this secret and thus is unable to forge a genuine-looking UI with
the correct secret. 

\item 
\bfdot{Malicious termination, modification or removal of \sol}
A simple but effective attack against \sol is terminating its kernel driver in
the OS, or even completely removing it. Similar to rootkit malware's termination
of host-based anti-malware defenses, rootkit ransomware can easily launch this
attack against \sol.\footnote{Malicious termination can be made difficult by registering \sol as a Windows Early Launch Antimalware (ELAM) driver.} The
pre-existing files on the protected partition remain immune to this attack; however, newly created or
updated files thereafter are not protected.  
As with the forged UI attack mitigation, the user-defined avatar or short phrase can 
indicate the correct execution of \sol in TXT/SVM, regardless of other tampering outside.
Minimal user diligence is needed to just make sure the 
\sol UI is regularly (automatically or manually) seen with this avatar or phrase.
Note that modifying the \sol updater's
binary, which may reside on the unprotected partition, does not help the
attacker; the SED unlock secret can only be accessed by the genuine \sol updater
(TPM unsealing).

\item
\bfdot{Known attacks against SEDs}
M\"{u}ller et al.~\cite{sed_attacks} show that SEDs are also vulnerable to known
attacks against software FDEs (e.g., cold boot, warm boot, DMA, and
evil-maid).  They also found a simple attack called \emph{hot plug}, 
enabled by the fact that SEDs are always in a binary state of locked or
unlocked. Once it becomes unlocked in a legitimate manner (e.g., user-supplied
unlock passwords), the adversary can connect the disk to another
attacker-controlled machine without cutting power, and can get access to
protected data. In addition to these attacks, an adversary may also capture the
cleartext SED secret/password from the SATA interface, e.g., by tapping the
connection pins with a logic analyzers.
Since all such attacks require physical access, i.e., desoldering
a microchip, manipulating the connector or evil maid attacks, they are not viable for a
scalable ransomware attack. 
More recently, different from the design limitations above, certain implementation
flaws have been identified by Meijer and van Gastel~\cite{sed-attacks}, which
severely affects SED security. To our understanding, the current SED-specific flaws,
e.g., user password and DEK not linked, mostly concern data confidentiality (with
physical access), whereas \sol's goal is data integrity (i.e., write protection).
However, there exist undocumented vendor-specific commands (VSCs)\footnote{These VSCs are like regular commands sent through the SATA or NVMe interface, which can be done by any privileged code.}
 on certain storage devices that allow flashing
unsigned firmware (directly or indirectly), which completely breaks \emph{TEE\textsubscript{Disk}}.
Among SED disks, Crucial (Micron) MX100 and MX200 are
vulnerable with such VSCs as reported by Meijer and van Gastel.
We may have to choose those SED drives without such VSCs
(e.g., after static analysis of firmware extracted via JTAG).
Caution is always necessary when taking devices with firmware as part of the TCB
(cf.\ the threat model of FlashGuard~\cite{flashguard}).
Last but not least, usually major SED manufacturers apply certain form of authentication/verification
of the firmware before being updated to the drive, e.g., Secure Downloads and Diagnostics.\footnote{Seagate (our SED) prevents counterfeit firmware: \url{https://www.seagate.com/ca/en/solutions/security/}}

\item \label{item:tpm-attack}
\bfdot{Attacks on TXT/TPM}
Although TPMs offer some physical tamper-resistance, TPMs and similar security
chips have been successfully attacked in the past (e.g.,~\cite{oslo, blackhat08,
dartmouth-tech, tpm-comm-attack1, tpm-comm-attack2}); see also Nemec et
al.~\cite{tpm-ccs17}.
However, with physical access excluded, we do not need to
consider these attacks; also note that tapping TPM pins and DMA attacks require
a malicious device to be connected.
Regarding known software-only attacks against TXT, most such attacks are
ad-hoc (e.g., the SINIT module flaw~\cite{another-txt}), or version-specific;
Intel has purportedly patched them in the subsequent versions, or at least the
user is motivated to choose one that has no known flaws.   \looseness=-1

There are also attacks
against TXT (e.g.,~\cite{attack-txt}) that exploit the System Management Mode
(SMM), an intrinsic part of the Intel x86 architecture, referred to as Ring -2. 
If the SMI (SMM interrupt) handler is compromised and SMI is left enabled, it can
preempt the TXT execution and intercept trusted operations. Although
no OS, hypervisor or bootloader runs in parallel with \sol to trigger SMI 
(e.g., by writing to port 0xB2), certain micro-architectural behavior may 
facilitate such attacks, e.g., the CPU temperature sensor. Nevertheless,
certain Intel CPUs leave SMI disabled after SINIT in TXT~\cite{mle}; we can also 
explicitly disable SMI generation with the Southbridge (model-specific) upon entry to our code, as SMI is not needed in \sol.
This at least significantly reduces the attack time window to just the number of CPU cycles needed to disable it.
Especially, for AMD CPUs~\cite{amd-svm-debug}, external SMI interrupts that assert after the start 
of SKINIT execution will be held pending (to ensure atomicity of SKINIT) until 
software subsequently sets GIF to 1 (internal SMIs are lost). We do not set GIF 
back to 1, and rely on polling instead (we do not gain extra performance with interrupts in our single-threaded environment).
Another possible (powerful) attack avenue similar to SMM is vulnerable Intel 
Management Engine firmware~\cite{me}. Unless there is a pressing need for ME, we suggest to 
disable it in a rigorous manner (for efforts and difficulties,
see~\cite{dis-me}, as there is no architecture support for disabling ME and SMI elegantly). 

\item \label{item:alt-dma}
\bfdot{Compromise-then-DMA attacks}
Although very unlikely to occur, we still consider a special situation where a DMA 
attack can be mounted but argue that it does not pose any threat. The remote
adversary or malware can compromise a programmable peripheral (e.g., a functionality-rich
gaming keyboard) and use it as the attack device. Then \sol's exposed DMA buffer
(not covered by Intel VT-d or AMD DEV protections) might be manipulated by that
compromised peripheral (malware/ransomware). Those buffers only contain content 
to be written, while LBA location/sector count is still sent via regular I/O (WRITE-DMA-EXT),
i.e., the adversary cannot point to the location of existing protected files.
Therefore, encrypting the DMA buffers buys the adversary no more than doing the same \emph{outside \sol}, 
i.e., contaminating files on the unprotected partition.
We can also reconfigure (every time inside TXT/SVM) DMA remapping in a way
that the concerned range is only accessible to a specific PCI device (SED). \looseness=-1

\item
\bfdot{Delayed attacks after deletion}
Persistent ransomware can stay hidden for a long period (ranging from weeks to
months), during which it just transparently decrypts encrypted data when
accessed~\cite{dormant}. This can trick the user to believe that her
data is intact (when viewed from within the OS). At some point, if she removes
older versions to save space or auto-deletion is triggered, then the ransom can 
be demanded (i.e., no more showing the decrypted version).

The root cause of this problem is that OS-based file viewers (e.g., Microsoft
Word), run outside the trusted environment and can be manipulated by rootkit
ransomware arbitrarily, such as performing decryption before displaying a file
to the user, or simply feeding a cached, unencrypted copy of the file.  A
straightforward countermeasure is to perform verification inside the updater
before removing previous versions, e.g., by porting advanced file viewing tools
in TXT/SVM, which can require significant effort. 
Such delayed attacks can be classified into two cases:
\begin{inparaenum}[1)]
\item Ransomware-triggered file updates to exhaust version limits/space.
\item Updates piggybacking on legitimate user edits.
\end{inparaenum}
Regular user files are either less frequently edited, or configured with a proper
schedule to consolidate frequent edits. Therefore, the latter would take a long
time for enough number of versions or age. We thus argue that the former would
be more effective for the adversary. Nevertheless, \sol can show an update
log (with the list of files committed) to the user in the trusted updater on each run.
It will raise an alert if the user notices files being committed that she has not touched.
According to our file selection principles (see Section~\ref{sec:del}), the number
of files being committed each time should account for a very small portion
of all the protected files (e.g., 5 files out of 1000) unless the total number is also small.

In general, if auto-deletion with aging is enabled, we suggest the duration 
should be long, e.g., a year or two, depending on the size of the protected partition.
Note that, delayed attacks risk being discovered and mitigated by anti-malware
vendors, and thus we do not consider them a serious threat. 

\item
\bfdot{Attacking auto-deletion with aging}
If older file versions are automatically deleted
after a preset threshold (e.g., 365 days), a straightforward threat is clock
source manipulation. Rootkit ransomware can adjust the system time
(to a far future date) to fool \sol to believe the versions are already too old to be kept.
To address this, \sol can be configured to only trust a signed NTP
time from a remote server, absence of which will stop auto-deletion (see Section~\ref{sec:del}).
\end{enumerate}
\section{Related work} \label{sec:rel}
There are many solutions dealing with user-level malware/ransomware for data protection; 
only FlashGuard~\cite{flashguard} targets rootkit-level ransomware. However, some solutions 
against data manipulation by rootkit malware
(not specific to rootkit-level ransomware) are close to \sol in spirit. We
discuss several examples from each category.

\subhead{Reverse-engineered keys}
Early-day ransomware had the (symmetric) file encryption keys embedded in their
obfuscated binaries, or stored in a C\&C server. Keys could be
recovered by reverse-engineering their code or intercepting C\&C traffic.
Ransomware now generally uses a public key to encrypt a
random file encryption key, and the private key remains only at the
attacker's machine (cf.~\cite{moti-96}), and thus much more resilient than
before; however, implementation flaws~\cite{flaw}
may still be leveraged to recover encryption keys. An exemplary
umbrella solution is \emph{NoMoreRansom}~\cite{no-ransom}, clustering file
recovery efforts from several public and industry partners. However, relying on
ransomware authors' mistakes is a non-solution, and finding such exploits may be too late for early victims.

\subhead{Offline and online backups}
An obvious recovery-based countermeasure against malware/ransomware is to 
make \emph{offline} backup of important data regularly (on media disconnected 
from the computer or with device-enforced write protection,
as ransomware also attempts to erase accessible backups).
Although simple in theory, effective deployment/use of backup tools could be
non-trivial, e.g., determining frequency of backups, checking integrity of
backups regularly (see Laszka et al.~\cite{economics} for an economic analysis
of paying ransom vs.\ backup strategies).  More problematically, the
disconnected/write-protected media must be connected/unlocked (online) during 
backup, at which point, malware/ransomware can encrypt/delete the files 
(see~\cite{flashguard, samas}). For cloud-based backup
systems, such as Dropbox (centralized) and \url{Syncthing.net} (P2P), a
potential issue is the size of their TCB (includes a full OS with multiple
network-facing servers), which may lead to large-scale data loss, if
compromised.

\subhead{Rootkit-level solutions}
S4~\cite{selfsecuring} is proposed as a self-securing storage entity behind a
security perimeter, which records all file operations (like journaling or
auditing) and retains old versions of user files. It is implemented as a network
service (similar to NFS), and assumed to be resistant to compromise by a remote
party (due to S4's limited outward interface).  The usage scenario is focused on
intrusion survival and forensics collection, in the case of an admin account
compromise in a client machine.  As S4 promptly stores all changes made to the
client machine, as soon as possible, its storage overhead can be significant.
To address this challenge, S4 makes use of novel compression and differential
versioning techniques, which can benefit \sol as well.  Also, 
without any TEE to ensure execution integrity and secrecy, it involves
the whole server infrastructure as the TCB, exposing many attack vectors.
More likely than a full system compromise, if the admin account of S4 (or any similar backup system) is
hacked, large volumes of data may be lost at once.

FlashGuard~\cite{flashguard} proposes to modify the garbage collection mechanism
of SSD firmware (assuming vendor support), so that for \emph{suspicious}
overwrites (i.e., first read and then written in a quick succession), a copy of
the original data block is kept for a preset amount of time (e.g., 20 days).
FlashGuard leverages a unique \emph{out-of-place write} feature of modern SSDs
(in contrast to regular hard drives), which provides an implicit backup of
recently overwritten data blocks. The user is expected to detect any attack
before the preset time elapses and perform the recovery from a separate machine;
otherwise the data will be lost.  The detection of suspicious overwrites can be
an issue; e.g., ransomware can read and encrypt the file, and at some later
point (i.e., not immediately to avoid being flagged), delete the file. However,
this can be solved by retaining \emph{all} deleted data blocks, at the expense of
increased storage overhead.  FlashGuard authors also do not specify the clock source to measure the preset time; SSDs do not offer any trusted clock, and relying on OS/BIOS could be fatal. 

Rootkit-resistant disks (RRD~\cite{rrd}) are designed to resist rootkit
infection of system binaries, which are labelled at installation time, and write
operations to protected binaries are mediated by the disk controller. System
binaries are updated by booting into a safe state in the presence of a security
token. While effective against rootkit infection, RRD is infeasible against
ransomware that targets regular user files (adding/updating will require
reboot). \sol's goals are complementary to RRD's and exclude protecting
system binaries.

\subhead{User-level solutions}
Defenses are usually
implemented as system services, kernel drivers (unprivileged adversary), or even user-land applications.
For instance, Redemption~\cite{redemption} explicitly mentions that their TCB
includes the display module, OS kernel, and underlying software.  Redemption
claims to provide real-time ransomware protection, by inspecting system-wide I/O
request patterns.  Its detection approach involves a comprehensive list of
features, with both content-based (entropy, overwriting and deletion) and
behavior-based (e.g., directory traversal).  In the end, a malice score is
calculated to facilitate decisions.  Redemption creates a protected area, called
\emph{reflected file}, which caches the write requests during inspection; the
file is periodically flushed to disk (if no anomaly is identified). This ensures data 
consistency in case of false positives, i.e., if the suspicious operations is confirmed
by the user to be benign, there is still the chance to restore the discarded data. \looseness=-1

In an effort to achieve better universality and robustness, some proposals
are purely data-centric (i.e., agnostic to program execution, checking
just the outcome). E.g., CryptoDrop~\cite{cryptodrop} focuses on file
transformation information for individual files, regardless of where those
transformations come from. It also claims to
achieve early detection.  It employs three novel indicators to detect suspicious file operations. Low file similarity before
and after may indicate encryption but legitimate operations can also cause it
(e.g., a blurred JPG file).  Shannon entropy can be used in detecting encryption
although compression also leads to high entropy. Last, file type changes
(through content parsing) might not be robust enough with format-preserving
encryption~\cite{fpe}.

Although most ransomware mitigation techniques aim to detect/prevent ransomware
as the primary goal, very few also focus on recovery, e.g.,
PayBreak~\cite{paybreak}. Symmetric keys used by ransomware to encrypt
user data are captured through crypto function hooking before they are 
encrypted with the adversary's public key, and then stored in a secure key vault. 
When infection is detected or a ransom is demanded, the user can retrieve
the keys for decryption without paying the ransom. PayBreak's crypto function 
hooking works for both statically and dynamically linked binaries, but only
if the ransomware uses known third-party crypto libraries. Also, it is subject to
evasion by obfuscation for statically linked ransomware. 
The key vault, even though encrypted with the user's public key and protected
by the admin privilege, can still be easily erased by rootkit ransomware.

ShieldFS~\cite{shieldfs} is a copy-on-write shadowing filesystem reactive to
ransomware detection, which is also based on I/O requests (I/O Request Packets - IRPs).
Its methodology fits in the intersection of recovery-based solutions and 
data loss prevention, and thus is similar to \sol in positioning.
The detection portion also makes use of numerous behavioral features reflected
from the IRPs. Specifically, ShieldFS's cryptographic primitives detection, 
different from PayBreak's, does not rely on hooking known crypto libraries, but 
captures inevitable properties of crypto primitives, such as the key schedule
pre-computation of block ciphers.
To achieve the claimed self-healing, on the first write attempt, ShieldFS keeps
a copy of the original file in a protected location (only from userland processes);
once an anomaly is detected, the changes made can be reverted with this copy,
or otherwise it can be deleted at any time. \looseness=-1

Microsoft BitLocker~\cite{bitlocker} is a widely-used (enterprise) data protection tool integrated with the Windows OS. BitLocker provides strong confidentiality guarantees 
through TPM-bound encryption. However, when a BitLocker-protected partition is unlocked after a successful boot (i.e., accessible
to the OS and applications), there is no way to distinguish a malicious write attempt 
from legitimate ones, and thus making the protected data vulnerable to even user-level ransomware attacks. \looseness=-1

 For advanced data protection in iOS, Apple's secure enclave co-processor (SEP~\cite{sep}) 
 is also a form of hardware security feature, enabling memory
 encryption and credentials management (among other functions).
 The SEP communicates with the application processors (APs) via a mechanism 
 called Secure Mailbox. 
 From the limited public documentation, it appears that per-application 
 access control is possible with SEP, therefore, decryption (and thus updates)
 can be only exposed to the right application.

Closest to \sol in design components is Pesos by Krahn et al.~\cite{pesos} but with different goals
and threat model; they use Intel SGX as \emph{TEE\textsubscript{Host}}
and Kinetic Open Storage as \emph{TEE\textsubscript{Disk}}. The high-level similarity is that \emph{TEE\textsubscript{Host}},
enforcing certain storage protection rules, connects to and is authenticated by \emph{TEE\textsubscript{Disk}},
and the Disk is capable of executing operations assigned by \emph{TEE\textsubscript{Host}}. The major contribution
is the flattened abstraction layers and a rich set of storage policies exposed.
Pesos assumes that only remote servers (the Host) are potentially malicious and the client machine 
is trusted, hence excluding (rootkit) ransomware/malware on the client machine. Therefore, 
trusted UI is no longer a problem, so the user can perform administrative operations (if needed)
from the client, such as specifying which files to delete through encrypted network to \emph{TEE\textsubscript{Host}}.
Since the kinetic storage supports secure communication (TLS), WAN network as an insecure Interface
can be used. In addition, if placed in \sol's setting, it suffers from the same issue of 
untrusted data source, i.e., there is no way to distinguish malicious writes from benign ones,
unless all applications can be ported and contained in SGX or TXT/SVM. \looseness=-1 

\section{Conclusions}

In summary, we propose to focus on data loss prevention, in an
effort to address rootkit-level data alteration as exemplified by ransomware, 
a significant threat that remains
largely unaddressed in current state-of-the-art solutions.
We leverage the trusted execution environments (TEEs) available with modern
computing devices and reason along various TEE placements between
the host and the storage device. Intel TXT and AMD SVM in conjunction with
TCG Opal SED are chosen for \sol as our current prototype.
\sol leaves original user files in use with applications and
exposes the protected copies as read-only all the time, and silently
accepts creation/modification of the files by preserving previous versions.
Users are only involved in file deletion occasionally in the trusted environment
(e.g., for regular file deletion or in case the protected partition becomes full).
Although our current prototypes are
less than ideal (e.g., file transfer performance), 
we believe \sol is a solid step towards countering rootkit ransomware. The source code of our prototypes will be made available through: \url{https://madiba.encs.concordia.ca/software.html}. \looseness=-1

\section*{Acknowledgment}

We are grateful to Jonathan M.\ McCune for guiding us in the final version of this paper. We also appreciate the help we received from the members of Concordia's Madiba Security Research Group. The second author is supported in part by an NSERC Discovery Grant.

\balance


\end{document}